\documentclass[superscriptaddress,preprint]{revtex4}
\usepackage{mathrsfs}
\usepackage{amssymb}
\usepackage[tbtags]{amsmath}
\usepackage{graphicx}
\usepackage{epsfig,graphicx,times}
\usepackage{color}
\usepackage{subfigure}
\usepackage{setspace}
\usepackage{epstopdf}
\usepackage{bm}
\usepackage[bookmarks=true,colorlinks,linkcolor=blue,urlcolor=blue,citecolor=blue,
plainpages=false,pdfpagelabels,final,breaklinks=true]{hyperref}
\begin{document}
\title{Sensitive detection of local magnetic field changes with atomic interferometry by using superconducting Meissner effects}
\author{Y. Q. Chai}
\affiliation{School of Physical Science and Technology, Southwest Jiaotong University, Chengdu 610031, China}
\affiliation{Information Quantum Technology Laboratory, International Cooperation Research Center of China Communication and Sensor Networks for Modern Transportation, School of Information Science and Technology, Southwest Jiaotong University, Chengdu 610031, China}

\author{M. Zhang\footnote{zhangmiao@swjtu.edu.cn}}
\affiliation{School of Physical Science and Technology, Southwest Jiaotong University, Chengdu 610031, China}

\author{L. F. Wei}
\affiliation{Information Quantum Technology Laboratory, International Cooperation Research Center of China Communication and Sensor Networks for Modern Transportation, School of Information Science and Technology, Southwest Jiaotong University, Chengdu 610031, China}

\date{\today}

\begin{abstract}
Sensitive detection of magnetic field is one of the open problem in metrology. Here, we propose an Mach-Zehnder atomic interferometry to sensitively detect the very weak local magnetic field, which is expelled by the superconductor (as the ``testing magnet") due to the Meissner effect. The induced magnetic field gradient near the superconductor provides a centripetal acceleration of the atomic motion in the interferometry and thus can be detected by using the atomic interferences. Given gravity acceleration of the atoms have been measured at the accuracy of $10^{-12}$ g, the measured sensitivity of the expelled local field could reach $10^{-14}$ T.
\end{abstract}

\maketitle
\section{introduction}
High precision measurements of physical quantities take the important roles in modern science and technology. As a powerful and precise measuring tool, atomic interferometry has been widely used for various metrologies~\cite{1,Zhu,3}, e.g., gyroscope~\cite{4}, gravimeter~\cite{5,6,7}, the measurement of fine structure constant~\cite{8}, the test of equivalence principle~\cite{EEP}, the demonstration of electrodynamics Aharonov-Casher (AC) effect~\cite{AC}, the gravitational Aharonov-Bohm (AB) effect~\cite{AB}, the gravitational wave detection~\cite{gravitational wave}, and the test of chameleon theories of dark energy~\cite{Dark Energy}. Particularly, the accuracy of measuring gravitational acceleration can reach $\delta g=10^{-12}g$~\cite{EEP} by using the Mach-Zehnder (MZ) atom interferometer. In this atom interferometer, three laser pulses are applied to split and recombine the atomic paths. The gravitational potential between the two separated paths of atom results in a signal $kgt^2$ of atomic matter-wave interference, with $k$ being the effective wave number of laser and $t$ the freely falling time of atom in vacuum. The high precision measurement on gravity is attributed to the large value of optical wave number $k$, the ensemble of  many cold atoms, and the atom-detection at single atom level.

It is well known that, in addition to the gravity field, there is also a magnetic field called usually as the geomagnetic field around the earth. The precise measurement of geomagnetic field has important significance and applications, such as for the geophysical exploration, geomagnetism, archaeology and so on~\cite{10,11,12,13}. Usually, the tool for measuring magnetic field is called as the magnetometer~\cite{14,15,16,17,18,19}. Given the superconducting quantum interferometer (SQUID) has been developed well for the various sensitive magnetometer applications~\cite{15}, and also the atomic interferometer possesses the high-precision measurement ability for the gravitational acceleration, in this paper we discuss how to combine these two metrology techniques for the implementation of the sensitivity detection of the earth's magnetic field. Basically, the earth's magnetic field is too weak to be detected directly by the atomic interferometer, which is originally designed to implement the sensitive measurement of  the gravitational acceleration. However, if a superconductor~\cite{Meissner0,Meissner1} served as the ``testing mass" is introduced, then the relevant earth's magnetic field is expelled outside (due to the Meissner effect), yielding a gradient of the geomagnetic field be generated. Furthermore, an atomic interferometer is designed to measure such a geomagnetic field gradient. As a consequence, the information of geomagnetic field in the local location of the conductor (which becomes the superconductor under the low temperature) can be extracted. Physically, the measurement accuracy of the local magnetic field is related to the one of the local gravitational acceleration.

This article is organized as follows. In Sec.~II, by specifically solving the Maxwell (London) equation outside (inside) the superconductor, we show how a geomagnetic field gradient is generated. In Sec.~III, an atomic interferometer is designed to implement the sensitive detection of such a magnetic gradient. Its achievable accuracy is also discussed. Finally, we
summarize our work in Sec. IV.

\section{The geomagnetic field gradient generated by using superconducting Meissner effect}
First, we discuss how to the local geomagnetic field gradient from the very weak geomagnetic field background. Let us introduce a cylindrical conductor to serve as the ``test mass", which can be cooled to become a superconductor. Before the cooling, the local geomagnetic field around it is assumed to be uniformly distributed and thus served as the very weak geomagnetic field background. By cooling the conductor to become superconductivity, the local background will be expelled due to the Meissner effect. The configuration is shown in Fig.~1.
\begin{figure}[htbp]
\includegraphics[width=6cm]{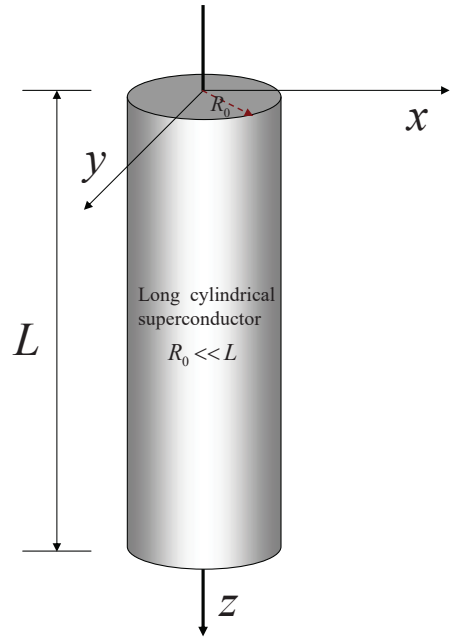}
\centering
\caption{The coordinate system is established with the axis of the superconducting cylinder as the $z$ axis. Considering the length $L$ of cylindrical superconductor is much larger than its radius $R_0$, then the magnetic field gradient along $z$-direction is negligible within the uniform  geomagnetic background.}
\end{figure}

\subsection{The magnetic field outside the superconductor}
Basically, the magnetic field $\textbf{B}$ outside the superconductor and its vector potential $\textbf{A}$ satisfies the relation:
\begin{equation}
\begin{aligned}
\nabla\times\textbf{B}=\nabla\times(\nabla\times\textbf{A})=-\nabla^2\textbf{A}=0\,,
\end{aligned}
\end{equation}
where $\nabla$ is the laplace operator and the Coulomb gauge $\nabla\cdot\textbf{A}=0$ is used. In the cylindrical coordinate system $(r,\theta,z)$ with $r=\sqrt{x^2+y^2}$ and $\theta=\arctan(y/x)$, the magnetic field reads
\begin{equation}
\begin{aligned}
\textbf{B}&=B_r(r,\theta)\textbf{e}_r
+B_\theta(r,\theta)\textbf{e}_\theta+B_z\textbf{e}_z\,,
\end{aligned}
\end{equation}
which can be further expressed as
\begin{equation}
\begin{aligned}
B_r(r,\theta)=\frac{1}{r}\frac{\partial A_z(r,\theta)}{\partial \theta}\,,
\end{aligned}
\end{equation}
\begin{equation}
\begin{aligned}
B_\theta(r,\theta)=-\frac{\partial A_z(r,\theta)}{\partial r}\,.
\end{aligned}
\end{equation}
Note that, the $B_z$ is the $z$-directional component of the magnetic field, which is invariant due to the symmetry of the long cylindrical superconductor. In Eqs.~(3) and (4), the $A_z$ is the $z$-directional component of the magnetic vector potential.

In the cylindrical coordinate system, the laplace equation (1) for $A_z$ outside the superconductor reads
\begin{equation}
\begin{aligned}
\frac{1}{r^2}\frac{\partial^2 A_z}{\partial\theta^2}+\frac{\partial^2 A_z}{\partial r^2}+\frac{1}{r}\frac{\partial A_z}{\partial r}=0\,.
\end{aligned}
\end{equation}
Writing $A_z=f(r)\Phi(\theta)$, the laplace equation becomes
\begin{equation}
\begin{aligned}
&r^2\frac{{\rm d}^2f}{{\rm d}r^2}+r\frac{{\rm d}f}{{\rm d}r}-\beta^2f=0\,,
\end{aligned}
\end{equation}
and
\begin{equation}
\begin{aligned}
&\frac{{\rm d}^2\Phi}{{\rm d}\theta^2}+\beta^2\Phi=0\,,
\end{aligned}
\end{equation}
with $\Phi(\theta)=\Phi(\theta+2\pi)$ and $\beta=0,1,2,\cdots$. The solutions of Eqs.~(6) and (7) are well-known
\begin{equation}
\begin{aligned}
f(r)=c_0+d_0\ln(r)+\sum_{n=1}^{\infty}\left(c_nr^n+d_nr^{-n}\right)\,,
\end{aligned}
\end{equation}
\begin{equation}
\begin{aligned}
\Phi(\theta)=\sum_{n=0}^{\infty}a_n\cos(n\theta)+b_n\sin(n\theta)\,,
\end{aligned}
\end{equation}
with $a_n$, $b_n$, $c_n$, $d_n$ being the arbitrary real numbers.

For $r\rightarrow\infty$, the magnetic field reduces to that of Earth, i.e.,
\begin{equation}
\begin{aligned}
\textbf{B}_0&=B_{0x}\textbf{e}_x
+B_{0y}\textbf{e}_y\\
&
=\left[B_{0x}\cos(\theta)+B_{0y}\sin(\theta)\right]\textbf{e}_r
+\left[B_{0y}\cos(\theta)-B_{0x}\sin(\theta)\right]\textbf{e}_\theta\,.
\end{aligned}
\end{equation}
This means that,
\begin{equation}
\left\{
\begin{aligned}
&\frac{1}{r}\frac{\partial A_z(r,\theta)}{\partial \theta}=B_{0x}\cos(\theta)+B_{0y}\sin(\theta)\,,\,\,\,\,{\rm with}\,\,r\rightarrow\infty\,,\\
\\
&-\frac{\partial A_z(r,\theta)}{\partial r}=B_{0y}\cos(\theta)-B_{0x}\sin(\theta)
\,,\,\,\,\,{\rm with}\,\,r\rightarrow\infty\,.
\end{aligned}
\right.
\end{equation}
According to Eqs.~(8), (9), and (11), we have
\begin{equation}
\begin{aligned}
A_z=C_0+r^{-1}\left[C_1\cos(\theta)+C_1\sin(\theta)\right]
+r\left[B_{0x}\sin(\theta)-B_{0y}\cos(\theta)\right]\,,
\end{aligned}
\end{equation}
and here $C_0$ and $C_1$ are the regularized coefficient.

\subsection{The magnetic field inside the superconductor}
The magnetic field $\tilde{\textbf{B}}$ inside the superconductor can be described by the London equation~\cite{London0,London1,London2}:
\begin{equation}
\begin{aligned}
\nabla\times\textbf{j}_s=-\frac{1}{\mu_0\lambda^2}\tilde{\textbf{B}}\,,
\end{aligned}
\end{equation}
with $\textbf{j}_s$, $\lambda$, and $\mu_0$ being the supercurrent density, London penetration depth, and the vacuum permeability, respectively. The above London equation can be further written as
\begin{equation}
\begin{aligned}
\nabla^2\tilde{\textbf{A}}-\frac{1}{\lambda^2}\tilde{\textbf{A}}=0\,,
\end{aligned}
\end{equation}
with $\textbf{j}_s=-\tilde{\textbf{A}}/\lambda^2$. Inside the superconductor, the magnetic potential $\tilde{A}_z=\tilde{f}(r)\tilde{\Phi}(\theta)$ is described by the London equation (14), i.e.,
\begin{equation}
\begin{aligned}
\frac{1}{r^2}\frac{\partial^2 \tilde{A}_z}{\partial\theta^2}+\frac{\partial^2 \tilde{A}_z}{\partial r^2}+\frac{1}{r}\frac{\partial \tilde{A}_z}{\partial r}=\frac{1}{\lambda^2}\tilde{A}_z\,.
\end{aligned}
\end{equation}
Obviously, the solution of $\tilde{\Phi}$ is similar to Eq.~(9), i.e.,
\begin{equation}
\begin{aligned}
\tilde{\Phi}(\theta)
=\sum_{n=0}^{\infty}\tilde{a}_n\cos(n\theta)+\tilde{b}_n\sin(n\theta)\,,
\end{aligned}
\end{equation}
with the arbitrary constant coefficients $\tilde{a}_n$ and $\tilde{b}_n$.
While, the function $\tilde{f}(r)$ obeys the following equation,
\begin{equation}
\begin{aligned}
r^2\frac{{\rm d}^2\tilde{f}}{{\rm d}r^2}+r\frac{{\rm d}\tilde{f}}{{\rm d}r}-\left(\frac{r^2}{\lambda^2}+n^2\right)\tilde{f}=0\,,
\end{aligned}
\end{equation}
which is solved as
\begin{equation}
\begin{aligned}
\tilde{f}(r)=\sum_{n=0}^{\infty}I_n\left(\frac{r}{\lambda}\right)\,.
\end{aligned}
\end{equation}
Here,
\begin{equation}
\begin{aligned}
I_n(x)=\left(\frac{x}{2}\right)^n
\sum_{k=0}^{\infty}\frac{1}{\Gamma(n+k+1)}\left(\frac{x}{2}\right)^{2k}\,,
\end{aligned}
\end{equation}
is the deformed Bessel function of order $n$. As consequence, the solution of $\tilde{A}_z$ reads
\begin{equation}
\begin{aligned}
\tilde{A}_z=\sum_{n,m}^{\infty}I_m\left(\frac{r}{\lambda}\right)
\left[\tilde{a}_n\cos(n\theta)+\tilde{b}_n\sin(n\theta)\right]\,.
\end{aligned}
\end{equation}

According the boundary condition $A_z(R_0)=\tilde{A}_z(R_0)$ at the surface of the cylindrical superconductor, we have
\begin{equation}
\begin{aligned}
\tilde{A}_z=I_0(\frac{r}{\lambda})\tilde{a}_0+I_1(\frac{r}{\lambda})
\left[\frac{B_{0x}(R_0+C_1R_0^{-1})}{I_1(\frac{R_0}{\lambda})}\sin(\theta)-
\frac{B_{0y}(R_0+C_1R_0^{-1})}{I_1(\frac{R_0}{\lambda})}\cos(\theta)\right]\,.
\end{aligned}
\end{equation}
Furthermore, using the continuity condition $\textbf{e}_\theta\cdot\textbf{B}(R_0)
=\textbf{e}_\theta\cdot\tilde{\textbf{B}}(R_0)$ of magnetic fields, i.e.,
\begin{equation}
\begin{aligned}
\frac{\partial A_z}{\partial r}\bigg|_{r=R_0}=\frac{\partial \tilde{A}_z}{\partial r}\bigg|_{r=R_0}\,,
\end{aligned}
\end{equation}
we have
\begin{equation}
\left\{
\begin{aligned}
&\tilde{a}_0=0\,,\\
\\
&C_1=2R_0\lambda\frac{I_1(\frac{R_0}{\lambda})}{I_0(\frac{R_0}{\lambda})}-R_0^2\,.
\end{aligned}
\right.
\end{equation}
According to Eqs.~(12), (21), and (23), the magnetic field outside and inside the superconductor can be obtained as,
\begin{equation}
\begin{aligned}
\textbf{B}=&\left[1-\frac{R_0^2}{r^2}+\frac{2R_0\lambda}{r^2}
\frac{I_1(\frac{R_0}{\lambda})}{I_0(\frac{R_0}{\lambda})}\right]
[B_{0y}\sin(\theta)+B_{0x}\cos(\theta)]\textbf{e}_r\\
&+\left[1+\frac{R_0^2}{r^2}-\frac{2R_0\lambda}{r^2}
\frac{I_1(\frac{R_0}{\lambda})}{I_0(\frac{R_0}{\lambda})}\right]
[B_{0y}\cos(\theta)-B_{0x}\sin(\theta)]\textbf{e}_\theta\\
=&B_x(x,y)\textbf{e}_x+B_y(x,y)\textbf{e}_y\,,
\end{aligned}
\end{equation}
and
\begin{equation}
\begin{aligned}
\tilde{\textbf{B}}=&\frac{2\lambda I_1(\frac{r}{\lambda})}{rI_0(\frac{R_0}{\lambda})}\left[ B_{0y}\sin(\theta)+B_{0x}\cos(\theta)\right]\textbf{e}_r\\
+&\frac{2}{I_0(\frac{R_0}{\lambda})}\left[I_0(\frac{r}{\lambda})-\frac{\lambda}{r}I_1(\frac{r}{\lambda})\right]
\left[B_{0y}\cos(\theta)-B_{0x}\sin(\theta)\right]
\textbf{e}_\theta\\
=&\tilde{B}_x(x,y)\textbf{e}_x+\tilde{B}_y(x,y)\textbf{e}_y\,.
\end{aligned}
\end{equation}
In the rectangular coordinate system, the magnetic field inside and outside the superconductor are expressed as
\begin{equation}
\left\{
\begin{aligned}
&B_x=B_{0x}+\frac{R_0(y^2-x^2)}{r^4}\left[R_0-2\lambda
\frac{I_1(\frac{R_0}{\lambda})}{I_0(\frac{R_0}{\lambda})}\right]B_{0x}
-\frac{R_0xy}{r^4}\left[R_0-2\lambda
\frac{I_1(\frac{R_0}{\lambda})}{I_0(\frac{R_0}{\lambda})} \right]B_{0y}\,,\\
\\
&B_y=B_{0y}+\frac{R_0(x^2-y^2)}{r^4}\left[R_0-2\lambda
\frac{I_1(\frac{R_0}{\lambda})}{I_0(\frac{R_0}{\lambda})}\right]B_{0y}
-\frac{R_0xy}{r^4}\left[R-2\lambda
\frac{I_1(\frac{R_0}{\lambda})}{I_0(\frac{R_0}{\lambda})} \right]B_{0x}\,,\\
\\
&\tilde{B}_x=\frac{2\lambda x^2}{r^3}
\frac{I_1(\frac{r}{\lambda})}{I_0(\frac{R_0}{\lambda})}
\left(\frac{y}{x}B_{0y}+B_{0x}\right)+\frac{2x^2}{r^2I_0(\frac{R_0}{\lambda})}
\left[I_0(\frac{r}{\lambda})-\frac{\lambda}{r}I_1(\frac{r}{\lambda})\right]
\left[\frac{y^2}{x^2}B_{0x}-\frac{y}{x}B_{0y}\right]\,,\\
\\
&\tilde{B}_y=\frac{2\lambda x^2}{r^3}\frac{I_1(\frac{r}{\lambda})}{I_0(\frac{R_0}{\lambda})}
\left(\frac{y^2}{x^2}B_{0y}+\frac{y}{x}B_{0x}\right)+\frac{2x^2}{r^2I_0(\frac{R_0}{\lambda})}
\left[I_0(\frac{r}{\lambda})-\frac{\lambda}{r}I_1(\frac{r}{\lambda})\right]
\left[B_{0y}-\frac{y}{x}B_{0x}\right]\,.
\end{aligned}
\right.
\end{equation}
Following Eqs.~(26), the Meissner effect of superconductor within the Earth's magnetic field is resolvable, as it showed in Fig.~2.

\begin{figure}[htbp]
\includegraphics[width=10cm]{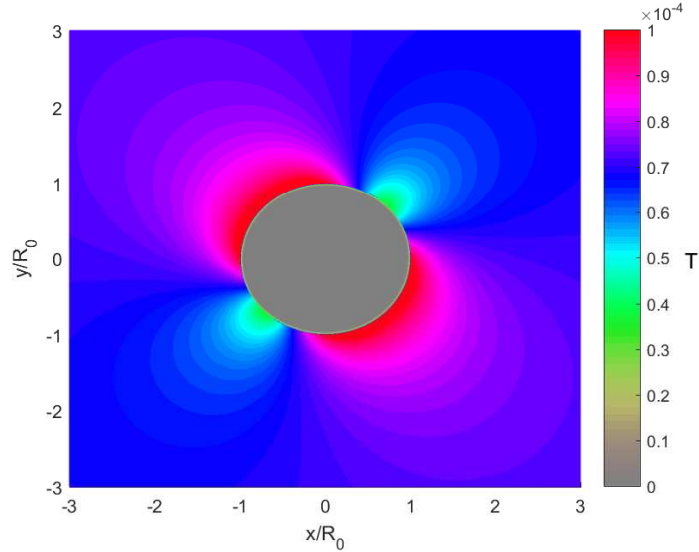}
\centering
\caption{Numerical solution of magnetic field~(26), with the radius $R_0=1$~cm of cylindrical superconductor, the London penetration depth $\lambda=0.01R_0$, and Earth's magnetic field $B_{0x}=B_{0y}=0.00005$~T. The figure shows the Meissner effect that the superconductor placed in an uniform magnetic field will affect the magnetic field distribution, resulting in a gradient magnetic field around the superconductor.}
\end{figure}

\section{Sensitive measurement of the generated geomagnetic field gradient by using the atomic interferometry}

\subsection{The motion of the cooled atom in a gradient magnetic field}
As shown in Sec.~II, the inhomogeneous magnetic field can be generated by the Meissner effect of superconductor in the uniform geomagnetic background. Such a gradient magnetic field influences the motion of the cooled free atoms with atomic internal magnetic dipoles.
The quantized Hamiltonian of atom reads
\begin{equation}
\begin{aligned}
\hat{H}&=\frac{\hat{\textbf{p}}^2}{2m}
+\bm{\mu}\cdot\textbf{B}(
\hat{x},\hat{y},\hat{z})+mg_z\hat{z}\,,
\end{aligned}
\end{equation}
with atomic kinetic energy $\hat{\textbf{p}}^2/(2m)$, magnetic moment $\bm{\mu}$, the gravitational acceleration $g=g_z\approx9.8~{\rm m/s^2}$ along $z$-direction.
To get the analytic solution of Hamiltonian (27), we consider the lowest order gradient of magnetic field (26).

For that, we denote the atomic horizontal position $(x=x_0+x',y=y_0+y')$, with $(x_0,y_0)$ being the initial position of atom, and $(x',y')$ the dynamically variable for horizontal motion.
Furthermore, considering the London penetration depth $\lambda\rightarrow0$, and the radius $R_0$ of cylindrical superconductor is much larger than $|x'|$ and $|y'|$, we can approximately write the magnetic field~(26) as follows by neglecting the high orders of $x'/R_0$ and $y'/R_0$,
\begin{equation}
\left\{
\begin{aligned}
B_x&\approx\left[1+\frac{(y_0^2-x_0^2)}{(x_0^2+y_0^2)^2}R_0^2
+\frac{x'}{R_0}\gamma_1
+\frac{y'}{R_0}\gamma_2\right]B_{0x}
-\left[\frac{x_0y_0}{(x_0^2+y_0^2)^2}R_0^2+\frac{x'}
{R_0}\gamma_3+\frac{y'}{R_0}\gamma_4 \right]B_{0y}\,,\\
\\
B_y&\approx\left[1+\frac{x_0^2-y_0^2}{(x_0^2+y_0^2)^2}R_0^2
-\frac{x'}{R_0}\gamma_1
-\frac{y'}{R_0}\gamma_2\right]B_{0y}
-\left[\frac{x_0y_0}{(x_0^2+y_0^2)^2}R_0^2+\frac{x'}
{R_0}\gamma_3+\frac{y'}{R_0}\gamma_4\right]B_{0x}\,.
\end{aligned}
\right.
\end{equation}

Here,
\begin{equation}
\left\{
\begin{aligned}
&\gamma_1=-\frac{2R_0^3x_0}{(x_0^2+y_0^2)^2}
\left[1+2\frac{y_0^2-x_0^2}{x_0^2+y_0^2}\right]\,,
\\
\\
&
\gamma_2=\frac{2R_0^3y_0}{(x_0^2+y_0^2)^2}
\left[1-2\frac{y_0^2-x_0^2}{x_0^2+y_0^2}\right]\,,
\\
\\
&\gamma_3=\frac{R_0^3y_0}{(x_0^2+y_0^2)^2}
\left[1-\frac{4x_0^2}{x_0^2+y_0^2}\right]\,,
\\
\\
&
\gamma_4=\frac{R_0^3x_0}{(x_0^2+y_0^2)^2}
\left[1-\frac{4y_0^2}{x_0^2+y_0^2}\right]\,.
\end{aligned}
\right.
\end{equation}

Following the magnetic field~(28), the Hamiltonian~(27) can be approximately written as
\begin{equation}
\begin{aligned}
\hat{H}&\approx\frac{\hat{\textbf{p}}^2}{2m}
+mg_x\hat{x}+mg_y\hat{y}+mg_z\hat{z}+{\rm constant}\,,
\end{aligned}
\end{equation}
with
\begin{equation}
\left\{
\begin{aligned}
g_x&=\frac{\mu_b}{mR_0}[(\gamma_1-\gamma_3)B_{0x}-(\gamma_1+\gamma_3)B_{0y}]\,,\\
\\
g_y&=\frac{\mu_b}{mR_0}[(\gamma_2-\gamma_4)B_{0x}-(\gamma_2+\gamma_4)B_{0y}]\,,
\end{aligned}
\right.
\end{equation}
being the horizontal  accelerations induced by the lowest order magnetic gradient, and where $\mu_b$ is the well-known Bohr magneton.
Immediately, we have Heisenberg operator of atomic position,
\begin{equation}
\begin{aligned}
\hat{\textbf{r}}(t)&=U^\dag(t)\hat{\textbf{r}}U(t)
=\hat{\textbf{r}}+\frac{\hat{\textbf{p}}}{m}t-\frac{1}{2}\textbf{g}t^2\,,
\end{aligned}
\end{equation}
with the well-known evolution operator $\hat{U}(t)=\exp(-it\hat{H}/\hbar)$ and the acceleration vector $\textbf{g}=(g_x,g_y,g_z)$.

Worth of note that, the Schr\"{o}dinger equation with Hamiltonian (30) is also solvable. For example, considering $x$-directional wave function of atom is initially cooled in the Gaussian one, i.e.,
\begin{equation}
\begin{aligned}
\langle x|\Psi(0)\rangle=\left(\frac{1}{2\pi\sigma_{x}^2}\right)^{-\frac{1}{4}}
\exp\left(-\frac{x^2}{4\sigma_{x}^2}\right)\,,
\end{aligned}
\end{equation}
then the time-dependent wave function reads
\begin{equation}
\begin{aligned}
\Psi(x,t)&=\langle x|\hat{U}(t)|\Psi(0)\rangle\\
&=\psi(t)
\exp\left[\frac{\hbar m}{\hbar m+2it\sigma_p^2}\left(\frac{4xg_xt^2-x^2-g_x^2t^4}{16\sigma_x^2}-im\frac{g_x^2t^3-2g_xt x}{2\hbar}\right)\right]\,,
\end{aligned}
\end{equation}
with
\begin{equation}
\begin{aligned}
\psi(t)=\left(\frac{1}{2\pi}\right)^{-\frac{1}{4}}\sqrt{\frac{2m\sigma_p}{\hbar m+2it\sigma_p^2}}\exp\left[-\frac{m^2g_x^2t^2}{4\sigma_p^2}-\frac{5img_x^2t^3}{6\hbar}
+\frac{\hbar m^3g_x^2t^2}{(\hbar m+2it\sigma_p^2)\sigma_p^2}\right]
\,.
\end{aligned}
\end{equation}
Above, $\sigma_p$ and $\sigma_x$ are atomic initial momentum and position uncertainties, respectively. The solutions of wave function $\Psi(y,t)$ and $\Psi(z,t)$ describing the $y$- and $z$-directional motions of atom have the similar forms as (34).

\subsection{The MZ atom interferometer}
Following Hamiltonian (30), we propose using the horizontal atom interference to measure the lowest order gradient magnetic field generated by Meissner effect within the background of Earth's magnetic field $\textbf{B}_0=\textbf{B}_{0x}\textbf{e}_x+\textbf{B}_{0y}\textbf{e}_y$.
\begin{figure}[htbp]
\includegraphics[width=8cm]{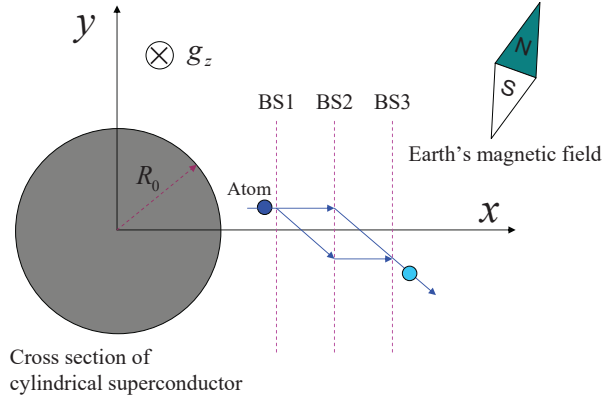}
\centering
\caption{Sketch for the MZ atom interferometer near the cross-section of cylindrical superconductor. Three laser pulses act as the atomic beam splittings, namely, BS1, BS2, and BS3, for realizing the matter-wave interference and measuring the magnetic fields gradient in $(x,y)$ plane generated by superconducting Meissner effect and the geomagnetic background. The gravitational acceleration of atom is along $z$ direction.}
\end{figure}
The MZ atom interferometer is dramatically showed by Fig.~3, where the atomic path is split and recombined by three short laser pulses, namely BS1, BS2, BS3. In experiments, the BS (atom-beam splitting) can be realized by the well-known Kapitza-Dirac (KD) scattering of atom in the pulsed standing wave of light~\cite{RMP-AKD,KD-E}. The light is large detuning with atomic internal states, generates the AC stark effect between two internal levels of atom, but not excite the transition between them. The AC stark effect is associated with the external state of atomic center-of-mass motion, and changes atomic momentum in terms of $\pm n2\hbar k$, with $k$ being the wave number of the laser beam, and $n=0,1,2,\cdots$ the order of KD scattering.
The probability of high-order KD scattering is small, so we consider just only the lowest-order KD scattering of atoms, i.e., $n=1$.
As a consequence, the separated distance between atom beams is on the order of  $(2\hbar k/m)\times t\approx6$~mm with $k=2\pi/(780~{\rm nm})$ and $t=0.5$~s.

The state of an atom moving along the two selected paths in Fig.~3 can be written as~\cite{27}
\begin{equation}
\begin{aligned}
|\psi_1\rangle=\hat{U}(t)e^{-i2\textbf{k}\cdot\textbf{r}}\hat{U}(t)
e^{i2\textbf{k}\cdot\textbf{r}}|\psi_0\rangle\,,
\end{aligned}
\end{equation}
and
\begin{equation}
\begin{aligned}
|\psi_2\rangle=e^{-i2\textbf{k}\cdot\textbf{r}}
\hat{U}(t)e^{i2\textbf{k}\cdot\textbf{r}}\hat{U}(t)|\psi_0\rangle\,.
\end{aligned}
\end{equation}
Here, $|\psi_0\rangle$ is atomic initial state, $\exp(\pm i2\textbf{k}\cdot\textbf{r})$ is the two-photon momentum recoil operator caused by the applied laser pulse, and $\hat{U}(t)$ the time evolution operator of atom between the laser pulses. The interference of atomic matter-wave is described by
\begin{equation}
\begin{aligned}
\langle\psi_1|\psi_2\rangle
&=\langle\psi_0|\hat{U}^{\dagger}(t)e^{-i2\textbf{k}\cdot\textbf{r}}
\hat{U}^{\dagger}(t)e^{i2\textbf{k}\cdot\textbf{r}}
\hat{U}(t)e^{-i2\textbf{k}\cdot\textbf{r}}\hat{U}(t)
e^{i2\textbf{k}\cdot\textbf{r}}|\psi_0\rangle\\
&=\langle\psi_0| e^{-i2\textbf{k}\cdot\textbf{r}(t)}e^{i2\textbf{k}\cdot\textbf{r}(2t)}
  e^{-i2\textbf{k}\cdot\textbf{r}(t)}e^{i2\textbf{k}\cdot\textbf{r}}|\psi_0\rangle\\
&=e^{i2\textbf{k}\cdot\textbf{g}t^2}\,.
\end{aligned}
\end{equation}
In the second line of this equation, we have used the
property $\hat{U}^{\dag}(t)\hat{U}(t)=\hat{U}(t)\hat{U}^{\dag}(t)=1$ of unitary evolution, and used the Heisenberg operator (32) of atom moving within the gradient magnetic field. For the third line of (38), we have used the so-called
Zassenhaus formula~\cite{28}
\begin{equation}
e^{\hat{A}+\hat{B}}=e^{\hat{A}}e^{\hat{B}}e^{-\frac{1}{2}[\hat{A},\hat{B}]}
e^{\frac{1}{6}[\hat{A},[\hat{A},\hat{B}]]+\frac{1}{3}[\hat{B},[\hat{A},\hat{B}]]}\cdots\,.
\end{equation}

The magnetic field induced phase-shift $2\textbf{k}\cdot\textbf{g}t^2$in Eq.~(38) has the form same as that of the standard MZ atom interferometer for gravitational acceleration measurements~\cite{Zhu}. Considering the applied laser beam is horizontal, i.e., $2\textbf{k}\cdot\textbf{g}t^2=2 (k_xg_x+k_yg_y)t^2$, we can directly use the experimental accuracy $\delta g_z$ of the gravitational acceleration detections to make an estimation on the accuracy of magnetic field measurements, for example, $\delta g_{x}=\delta g_{y}=\delta g_z\approx10^{-8}~{\rm m/s^2}$ with the freely falling time $t=160$~ms~\cite{Zhu}.
According to magnetic field induced acceleration (31), we have
\begin{equation}
\left\{
\begin{aligned}
\delta B_{0x}&=\frac{mR_0\delta g_x}{\mu_b(\gamma_1-\gamma_3)}+\frac{mR_0(\gamma_1+\gamma_3)[\delta g_y(\gamma_1-\gamma_3)-\delta g_x(\gamma_2-\gamma_4)]}{2\mu_b(\gamma_2\gamma_3-\gamma_1\gamma_4)
(\gamma_1-\gamma_3)}\,,
\\
\\
\delta B_{0y}&=\frac{mR_0[\delta g_y(\gamma_1-\gamma_3)-\delta g_x(\gamma_2-\gamma_4)]}{2\mu_b(\gamma_2\gamma_3-\gamma_1\gamma_4)}\,.
\end{aligned}
\right.
\end{equation}
Numerically, considering the initial position $(x_0,y_0)=(2,2)$~cm of atom, then $\gamma_1=-\gamma_2=\gamma_3=\gamma_4\approx0.06$, with atomic mass $m=10^{-25}$~kg and the radius $R_0=1$~cm of cylindrical superconductor. As a consequence, the measuring accuracy for Earth's magnetic field is on the order of $B_{0x}\approx B_{0y}\approx0.00005\pm10^{-11}$~T. Note that, the measurement-accuracy of atomic acceleration can be significantly improved by increasing the freely falling time $t$. For example, the experiment~\cite{EEP} showed the relative acceleration between the isotope atoms at the level of $10^{-12}g_z$. This means that the accuracy of geomagnetic field measurements may reach to a level of
$B_{0x}\approx B_{0y}\approx0.00005\pm10^{-14}$~T. In the ground-based laboratory, the freely falling time of atoms is ultimately limited by the sizes of practical vacuum installation. There are already plans for testing the matter-wave interference by launching the space-mission~\cite{29}.
The advantages of space experiments are the long freely falling time of test particles and the small nongravity noises. Thus, the future measurements for magnetic fields could reach at the unprecedented accuracy by using atomic matter-wave interferometry.

\section{conclusion}
We have showed an approach to measure the local magnetic field changes by using atomic matter-wave interferometry and the superconducting Meissner effects.
Firstly, we computed the magnetic field distributions inside and outside a cylindrical superconductor within the geomagnetic field background. Secondly, we designed a horizontal MZ atom interferometer to measure the nonuniform magnetic field generated by the superconducting Meissner effects and the uniform  geomagnetic field background. This atom interferometer is realized by three laser pulses which act as the atomic beam splittings but do not excite the atomic internal states. We considered the lowest order magnetic gradient and showed that the induced phase shift has the form same as that of the standard MZ atom interferometer for measuring the gravitational acceleration.
Compared with the standard MZ atom interferometer, the proposed measurement does not significantly change the experimental installation, just place a superconductor near the path of freely falling atoms and use the horizontal laser pulses. The scheme relies on the Meissner effect of superconductors, so it is not only suitable for measuring the geomagnetic field, but also suitable for measuring any magnetic field and even the superconductivity itself such as the London penetration depth.

\end{document}